# Convolution Inference via Synchronization of a Coupled CMOS Oscillator Array

D. E. Nikonov[1], P. Kurahashi[2], J. S. Ayers[2], H.-J. Lee[2], Y. Fan[2], and I. A. Young[1]
[1]Components Research, TMG, Intel, Hillsboro, USA, email: dmitri.e.nikonov@intel.com
[2]Advanced Design, TMG, Intel, Hillsboro, USA

*Abstract* — Oscillator neural networks (ONN) are a promising hardware option for artificial intelligence. With an abundance of theoretical treatments of ONNs, few experimental implementations exist to date. In contrast to prior publications of only building block functionality, we report a practical experimental demonstration of neural computing using an ONN. The arrays contain 26 CMOS ring oscillators in the GHz range of frequencies tuned by image data and filters. Synchronization of oscillators results in an analog output voltage approximating convolution neural network operation.

## I. WHY OSCILLATOR NEURAL NETWORKS?

The demand for artificial intelligence and machine learning computing has experienced an explosive growth in recent years. While most of this computing is performed by traditional digital hardware, the share of specialized hardware based on neural networks (NN), is growing. An option for NN, the oscillator neural networks (ONNs) [1] was found to be a promising approach when benchmarking [2] against the artificial NN and cellular NN [2].

While a considerable amount of theoretical and simulation work on ONN has been previously published, the majority of prior work focuses on implementing associative memory using the phase shift keying (PSK) approach [3]. Fabricated CMOS oscillator chips [4] show functional circuits, but not the actual neural computing. "Beyond-CMOS" device oscillator arrays, e.g. with spin-torque oscillators, have faced limitations due to their ability to synchronize only a few of the oscillators [5]. In this work we demonstrate neural processing, specifically convolution inference, performed by the more robust frequency shift keying (FSK) of oscillators in an array with a sufficiently large number of oscillators.

## II. CONVOLUTION AND GABOR FILTERS

One of the ubiquitous algorithms in image recognition for neural processing is a convolutional neural network (CoNN). The most compute intensive operation in a CoNN is a convolution of an image with a set of filters/kernels. An example 'Lenna' image and its fragment, $F$, of 5x5 pixels is shown in Fig. 1. Here we use a set of Gabor filters, $G$, as shown in Fig. 2, where oscillating patterns with various orientation and various spatial periods of oscillations are indicated. Pixel convolution of the above fragments and filters is given by the expression

$$(F \cdot G)_{kl} = \sum_{i,j=1}^{n} F_{k-i,l-j} G_{ij}, \quad (1)$$

which is equivalent to a dot product of corresponding real-numbered vectors. On the other hand, the dot product is related to their Euclidean distance and absolute values as:

$$2(F \cdot G) = |F|^2 + |G|^2 - |F - G|^2. \quad (2)$$

In Fig. 2 four of the filters which have the largest dot product i.e., the smallest distance to the fragment, are highlighted.

## III. SYNCHRONIZATION AND DEGREE OF MATCH

The general features of oscillator synchronization can be studied using a model of non-linear oscillators similar to a complex-numbered van der Pol equation [3] $\omega_i$ and a coupling coefficient $\varepsilon$:

$$\frac{dz_i}{dt} = (\rho + i\omega_i)z_i - \rho z_i |z_i|^2 + \varepsilon \sum_{j=1}^{n} z_j. \quad (3)$$

Two coupled oscillators synchronize and lock their phases only if their initial detuning (difference of frequencies) are within the locking range determined by the strength of coupling.

$$|\Delta \omega| = |\omega_2 - \omega_1| < \varepsilon. \quad (4)$$

An array of oscillators, Fig. 5, needs to have as many oscillators, $n$, as there are pixels in the fragment. The most efficient way to couple the oscillators is through a common node, "averager", which sums the oscillator signal outputs and broadcasts the sum signal back to their inputs. The fragment and a filter are fed into an array via FSK of initial frequencies [6]:

$$\omega_j = \omega_0 + \Delta\omega(F_j - G_j). \quad (5)$$

The Degree of Match (DOM) is equal to the magnitude of the analog output envelope of the AC signal at the averager. The behavior varies depending on the amount of synchronization of the oscillators, i.e. value of the dot product. For a *good match, Fig. 3*, frequencies synchronize close to the center one. Relative phases lock to certain values, and the DOM signal grows and stays constantly high. For a b*ad match, Fig. 4*, frequencies remain spread out. Relative phases grow with time, and the DOM signal stays low, strongly oscillating with no oscillator synchronization (i.e., experiences beats). The final frequencies for various filters are shown in Fig. 7. This results in an empirical relationship between the dot product and the

DOM signal, in Fig. 6. The locking time was found to require the number of periods according to:

$$T_{lock} \approx 2.4\omega_0 / \Delta\omega \approx 1.2\omega_0 / (\varepsilon n). \qquad (6)$$

## IV. ARRAY OF CMOS RING OSCILLATORS

The coupled CMOS oscillator array (COCOA) test chip was implemented following the design of [7] using the Intel 22nm process [8]. Ring oscillators with 3, 5, 7 inverter stages were fabricated and found to have qualitatively similar performance. Each array, Fig. 8, includes: 26 ring oscillators connected to the averager with variable capacitors, and using 26 current based digital to analog converters (DAC) to shift oscillator frequencies, and a peak detector which approximates the envelope of the averager to extract the DOM signal. The 26-oscillator array with all auxiliary circuits occupies an area of 88μm x 89μm.

The setup, Fig. 9, generates digital codes for the DACs to shift the oscillator frequencies, provides a few analog control voltages and currents, and detects the analog time-dependent voltage at the output. Measured oscillator frequencies range up to 8GHz for 3-stage, 6GHz for 5-stage, and 4GHz for 7-stage oscillators, in agreement with circuit simulations. This ensures fast inference times. Oscillators synchronize if their frequencies are within the locking range, related to the coupling capacitance $C_{coup}$. For an amplitude of driving current $I_{drv}=0.26mA$, supply voltage $V_{cc}=0.8V$, $f=6GHz$, and $C_{coup}=1fF$, the locking range is

$$\frac{\Delta\omega}{\omega} < \frac{2\pi f C V_{cc}}{I_{drv}} \sim 0.14. \qquad (7)$$

The optimum value of $C_{coup}$ is chosen with the requirement that oscillator frequencies are within the locking range for a good match, while they are mostly outside the locking range for a bad match. The power per oscillator is approximately

$$P_{osc} = I_{drv} V_{cc} \sim 0.2 mW. \qquad (8)$$

## V. DEGREE OF MATCH AND DEMONSTRATED INFERENCE

The measurements are consecutively run for each of the Gabor filters on a given fragment of the image. The peak detector output approximates the maximum envelope of the AC signal at the averager. The resulting output waveforms from the peak detector (for the 5-stage oscillators) are shown in Fig. 10. The oscillations of the oscillators are triggered by a fast enable voltage at 0ns on the plot scale, the peak detector output enable is delayed until 9ns later. We sample the peak detector voltage at 15ns and use it as the DOM. Based on 6ns delay per convolution and the above power, we benchmark that the present ONN has ~10x shorter delay per inference and ~5x lower energy per inference in a CoNN compared to typical neural accelerators such as Google TPU or Eyeriss [2].

The relationship between the dot product and the DOM is shown in Fig. 11 for the 5-stage oscillators. Similar results are observed for oscillators with a different number of stages. The dynamic range is defined as the difference in the DOM between the best and worst matched filters. It is important for the subsequent winner-take-all operation.

The correlation between the dot product and DOM is close to a linear correlation. However the relationship is non-monotonic due to non-monotonic evolution of frequencies under the influence of the rest of the oscillators and due to arbitrary initial phases of oscillators. Simulated and measured DOM are in good agreement.

In conclusion, we have experimentally demonstrated that the calculation of convolution can be approximated by the analog output signal, DOM, for arrays of coupled oscillators. This demonstration of coupled oscillator synchronization behavior opens the possibility for other emerging nanoscale oscillator technologies to enable more energy efficient accelerators for artificial intelligence.

### ACKNOWLEDGMENT

The authors gratefully acknowledge the contribution of their Intel colleagues Abhishek Tiwari, Telesphor Kamgaing, George Dogiamis, Aleks Aleksov, and Johanna Swan.

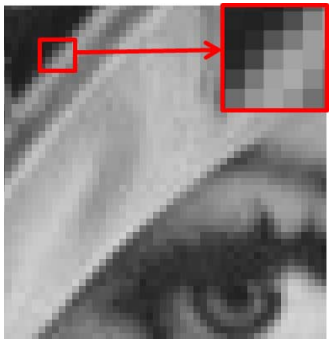

Fig. 1. An example 5x5-pixel fragment of the Lenna image used in the convolution.

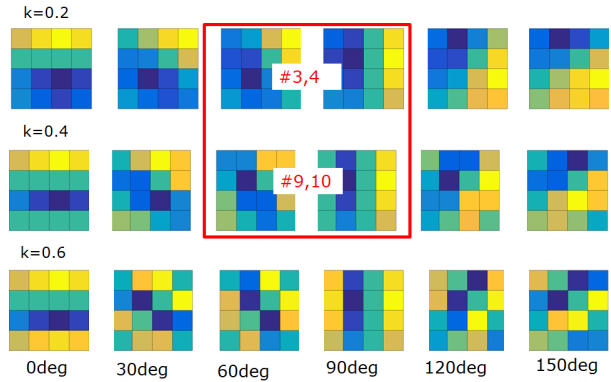

Fig. 2. A set of Gabor filters used for the convolution. Yellow color corresponds to +1, blue color to -1. Gabor filters are oscillatory patterns. The direction (in degrees) and the inverse spatial period of oscillations, $k$, indicated..

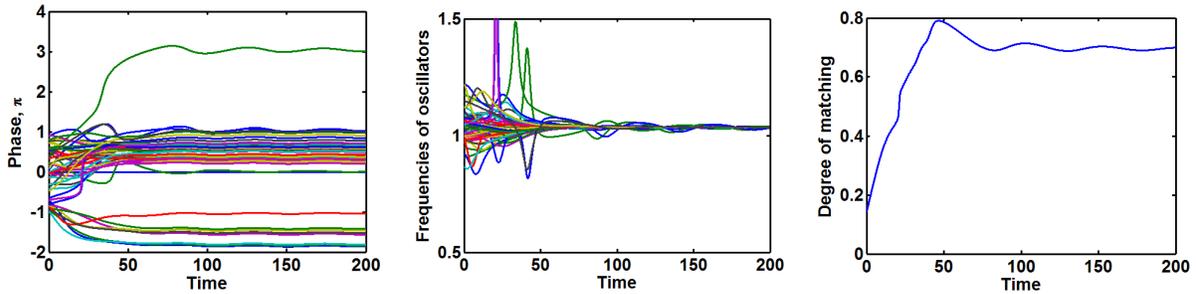

Fig. 3. Simulated phase, frequency of oscillators and the envelope of the waveform at the averager for a case of a good match (filter #3). The time is in units of radians, i.e. the period is $2\pi$. The frequency is in units of center frequency $\omega_0$. DOM is dimensionless here.

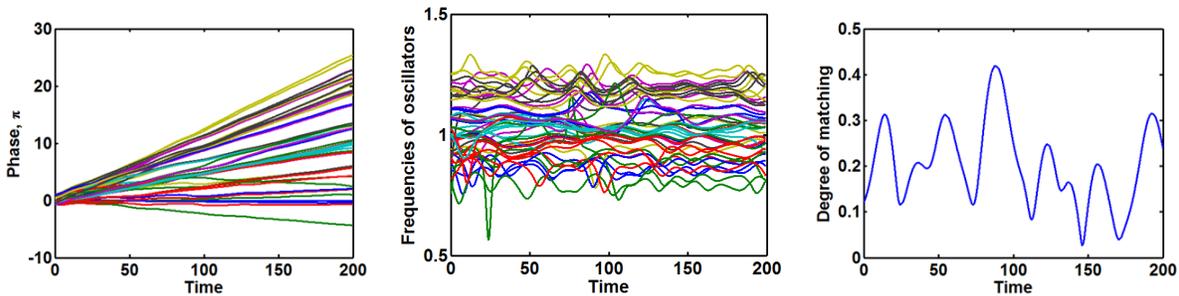

Fig. 4. Simulated phase, frequency of oscillators and the envelope of the waveform at the averager for a case of a bad match (filter #18).

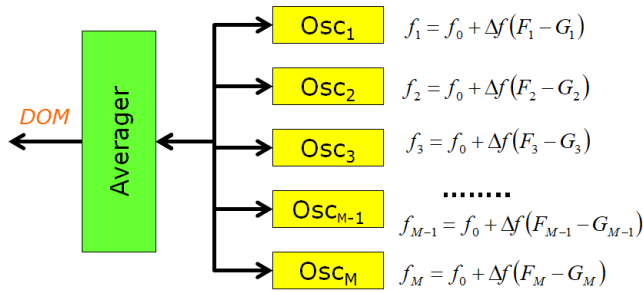

Fig. 5. Block diagram of a coupled oscillator array with FSK. Signals are passed in both directions through links.

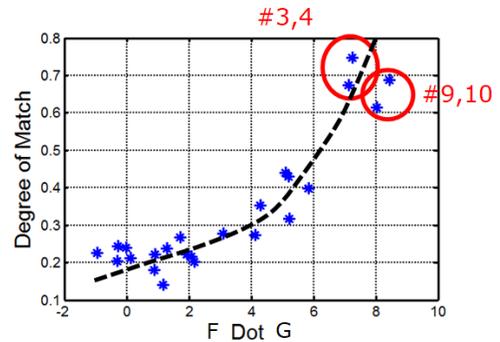

Fig. 6. Correspondence of the degree of match and the dot product of the fragment F and 18 Gabor filters G.

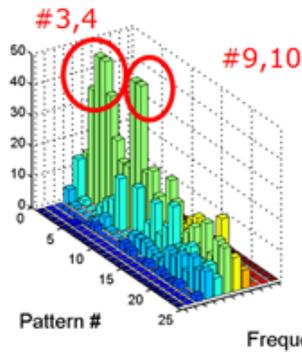

Fig. 7. Spectra of final frequencies of oscillators in the array for various Gabor filters.

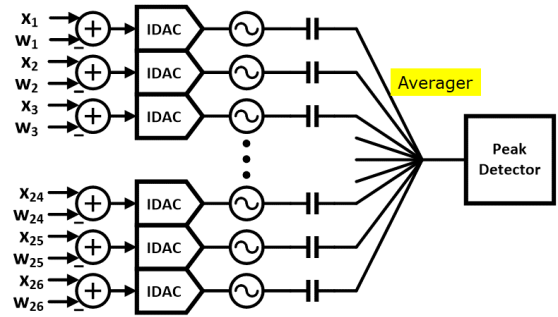

Fig. 8. Block diagram of the CMOS ring oscillator test chip.

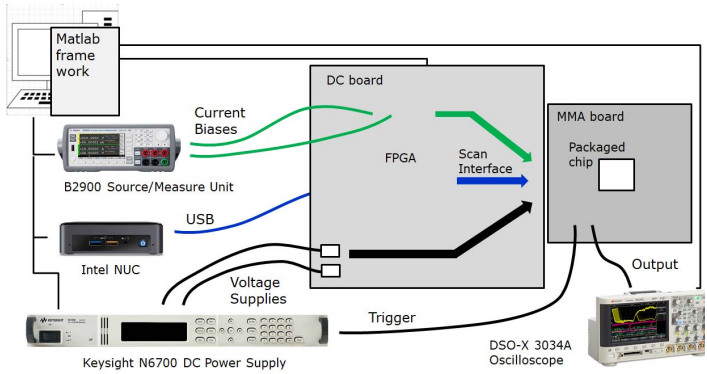

Fig. 9. Experimental setup for measuring of the DOM voltage.

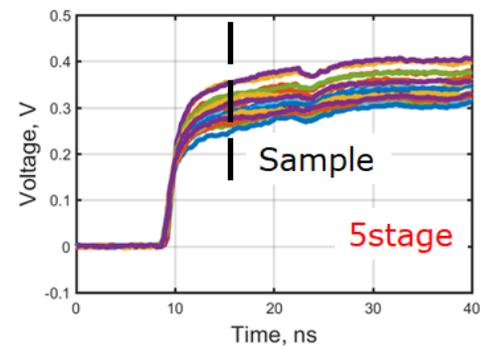

Fig. 10. Measured waveforms at the output of the peak detector in the 5-stage oscillator array.

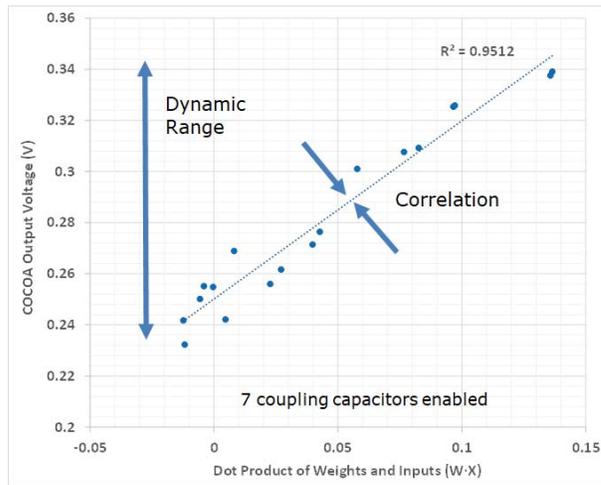
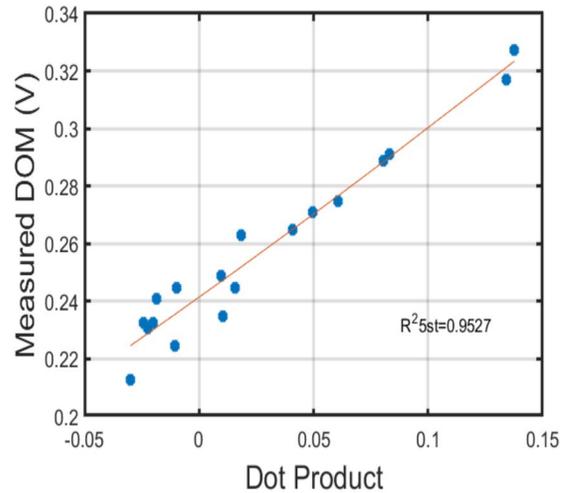

Fig. 11. Comparison of the simulated and measured relations between the DOM voltage in the 5-stage oscillator array and the dot product of the 18 Gabor filters and an image fragment.